# Measurement of dynamic light scattering intensity in gels

Cyrille Rochas[1] and Erik Geissler[2,3*]

[1] Univ. Grenoble Alpes, CNRS, CERMAV, F-38000 Grenoble, France

[2] Univ. Grenoble Alpes, LIPhy, F-38000 Grenoble, France

[3] CNRS, LIPhy, F-38000 Grenoble, France

* erik.geissler@ujf-grenoble.fr

**Abstract**

In the scientific literature little attention has been given to the use of dynamic light scattering (DLS) as a tool for extracting the thermodynamic information contained in the absolute intensity of light scattered by gels. In this article we show that DLS yields reliable measurements of the intensity of light scattered by the thermodynamic fluctuations, not only in aqueous polymer solutions, but also in hydrogels. In hydrogels, light scattered by osmotic fluctuations is heterodyned by that from static or slowly varying inhomogeneities. The two components are separable owing to their different time scales, giving good experimental agreement with macroscopic measurements of the osmotic pressure. DLS measurements in gels are, however, tributary to depolarised light scattering from the network as well as to multiple light scattering. The paper examines these effects, as well as the instrumental corrections required to determine the osmotic modulus. For guest polymers trapped in a hydrogel the measured intensity, extrapolated to zero concentration, is identical to that found by static light scattering from the same polymers in solution. The gel environment modifies the second and third virial coefficients, providing a means of evaluating the interaction between the polymers and the gel.

**Introduction**

In the arsenal of techniques for characterising polymers and suspensions of solids, dynamic light scattering (DLS) is one of the more powerful weapons [1-3]. Although its most common use may be for determining the hydrodynamic radius $R_H$ of particles in suspension, its major strength lies in its ability to measure molecular weights and interactions of polymers in solution. DLS also possesses a long pedigree in the field of polymer gels and polymer solutions [4-10]. Here too, however, most reports have focussed on the hydrodynamic correlation length $\xi_H$ rather than on the thermodynamic information enclosed in the intensity of the scattered light. This could in part be because the calibrations required for intensity measurements are more exacting, but it is equally likely that the underlying reason is doubt about the reliability of such measurements. The purpose of this article is to show that the DLS method can indeed provide robust and reproducible measurements of the intensity of light scattered from gels.

Here we examine in detail some of the practical aspects of DLS in hydrogels, and some precautions that should be taken. We report measurements in a wide range of heterodyning conditions on gels that are either composed of or contain fluctuating polymers, for which these considerations may affect the precision of the results. It should however be pointed out that in the systems investigated here, the solvent is water: the small Rayleigh ratio of water simplifies the analysis of the high frequency density fluctuations in the solvent.

**Theoretical background**

The background formalism of DLS is well established[1-10] and is outlined here merely to place the discussion in context. In a wide variety of polymer gels and polymer solutions the light scattered by the fast fluctuations of the polymer chains $I_{dyn}(q)$ is accompanied by a more intense component scattered by large clusters or quasi-static inhomogeneities that are associated with the elastic strains in the network. Owing to their very different relaxation rates, these fast and slow components are generally easy to distinguish in the time correlation function of the intensity $G(q,\tau)$ that is measured in a DLS experiment, where

$$G(q,\tau) = <I(q,t)I(q,t+\tau)>/<I(q,t)>^2 \qquad (1)$$

In this expression, $I(q,t)$ is the instantaneous intensity measured at a given angle $\theta$ by the detector at time $t$, and $I(q,t+\tau)$ that measured at a later time $t+\tau$. The brackets $<>$ signify an average taken of the whole experimental duration, and the transfer wave vector $q$ is equal to $(4\pi n/\lambda)\sin(\theta/2)$, where $n$ is the refractive index of the medium and $\lambda$ is the wavelength of the incident light. Since the fast and the slow components of the light originate from the same region of the sample, they arrive in phase at the detector and the resultant optical heterodyning enables quantitative estimates to be made of the mean intensity $I_{dyn}(q)$ of the fast dynamic component. This is the component that contains the information on the osmotic properties of the gel, through the relationship

$$I_{dyn}(0) = \frac{KkTc}{\partial \Pi / \partial c} \qquad (2)$$

where $I_{dyn}(0)$ refers to the mean intensity of the dynamically scattered light, extrapolated to zero angle, $k$ is Boltzmann's constant, $T$ the absolute temperature, $\Pi$ the osmotic pressure, and $c$ is the polymer concentration. In Eq. 2, $I_{dyn}(0)$ is normalised with respect to a known standard, generally toluene. The optical contrast factor $K$ between the polymer and solvent, in the case of vertically polarised light detected in the plane perpendicular to the polarisation axis, is defined by

$$K = (2\pi n_0 dn/dc)^2 / \lambda^4 \qquad (3)$$

where $dn/dc$ is the refractive index increment between polymer and solvent, and $n_0$ is the refractive index of the liquid in the index matching bath (again, usually toluene) [11].

The slow component, by contrast, is not necessarily of thermodynamic origin. In gels, for example, mechanical vibrations transmitted by cooling pumps or even from machinery in other parts of the building often excite resonances in the optical table that give rise to an oscillatory component in the intensity correlation function of Eq. 1. Small changes in temperature also cause slow rearrangements in the gel that appear as a slow mode. In concentrated polymer solutions, intense slow modes are a common feature, caused by large clusters that diffuse and also exhibit slow internal fluctuations. Owing to the inverse relationship between intensity and osmotic pressure, however, their large scattering intensity signifies that they contribute little, or even nothing, to the osmotic pressure of the solution.

It is usually a straightforward matter to discriminate the fast from the slow component in the intensity correlation function. This allows estimates to be made of the osmotic modulus $\kappa = c\partial\Pi/\partial c$ that are in excellent agreement with independent osmotic pressure measurements [8,12]. Observations by DLS of mobile molecules trapped in rigid hydrogels have similarly been used to determine the molecular weight of the guest molecules.[13,14] Compared to measurements of macromolecules in free solution, trapping by a gel can be advantageous, since the gel network immobilizes the large aggregates and dust particles, which are difficult to remove and often perturb measurements in conventional polymer solutions. In these earlier studies, the scattered intensity $I_{dyn}(q)$ of the mobile polymer in the gel and its diffusion coefficient $D$ were found to be related to the osmotic pressure $\Pi$ in the same way as in the free solution.

**Experimental**

The DLS instrument consisted of an ALV/LSE5004 goniometer and an ALV 7004 digital correlator equipped with near-monomode optical fibre coupling and pseudo-cross correlation. The light source was a 25 mW HeNe laser working at 632.8 nm. For the gels, measurements were made at 30, 50, 70, 90, 115 and 150° for a given sample, after which the position of the tube was changed manually and the measurements repeated at each of the 6 angles. For each sample, between 6 and 9 positions were examined. In this paper, however, we focus only on the intensity measurements, rather than on the angular variation of the scattered light, which was discussed elsewhere.[14]

The polyacrylamide gels were prepared in the standard way[15] by dissolving the required amount of acrylamide (Acros organics) together with N,N'-dihydroxyethylene-bis-acrylamide in a weight ratio of 50:1. Ammonium persulfate was used to initiate the polymerisation, and TEMED was added to make the pH of the precursor solution basic. The solutions were poured into cylindrical glass tubes and allowed to polymerise at room temperature. Gelation took place within half an hour, and the samples were left for one week for the reaction to complete.

Dextran, of three nominal molecular weights $M_w$=4.6×10$^5$, 5.0×10$^5$ and 2×10$^6$ Daltons, supplied by Sigma, were prepared at concentrations 1, 3, 5, 7.5, 10, and 15 g/l, either in distilled water alone, or, for incorporation into hydrogels, in aqueous solution with agarose (Hispanagar, Burgos, Spain) at concentration 5 or 10 g/l.[16, 17] These gel precursor solutions were heated to 100ºC with stirring and then allowed to cool in 10 mm diameter cylindrical glass tubes. As the concentration range explored in these measurements lay outside the region of phase separation,[14] the optical density of the agarose gels remained close to zero. To check that imperfections in the glass tubes were not a source of error, measurements were also made with precision cylindrical quartz cells (Hellma), as well as in 3 mm glass tubes.

The transmission factors of the samples were determined by placing the cylindrical sample tubes in rectangular quartz optical cells, with water filling the intervening space. The measurements were performed on a Varian Cary 50 Bio spectrophotometer set at 632.8 nm.

**Results and discussion**

*Ergodic regime*

In most circumstances, the intensity of the light $I(q,t)$ scattered by a polymer solution, or a suspension with a large number of freely mobile particles, obeys Gaussian statistics. The *intensity* correlation function in Eq. 1 is then governed by the Siegert relation[18]

$$G(q,\tau) = 1 + \beta|g(q,\tau)|^2 \qquad (4)$$

where $g(q,\tau)$ is the *field* correlation function with $|g(q,0)|=1$. In what follows, we write for the total scattered intensity $<I(q,t)>=I$, where both the time average and the transfer momentum $q$ are implicit. The optical coherence factor $\beta$ in Eq. 4, which is characteristic of the instrument, is determined by the angle subtended by the pinholes or by the optical fibre that transmits light into the detector. Ideally, $\beta=1$. Its true value can be estimated by analysing the light scattered by a dilute suspension of particles in a weakly scattering medium. In this work, the response of the DLS instrument was determined by measuring the light scattered from aqueous suspensions of polystyrene latex (Dow Chemical).

Eq. 4 is the response of the correlator when the detection system operates in a linear regime. In practice, account should be taken of the dead time $\tau_d$ of the detector and its noise $b$. The extent to which the coherence factor depends on these instrumental parameters is displayed as a function of the total detector count rate $I$ in **Figure S1** of the Supporting Information. Correction for these instrumental effects requires that the apparent value of $\beta$ be divided by

$$f(I) = [(1-\tau_d I)(1-b/I)]^2 \qquad (5)$$

*Gels*

Polymer gels consist of a matrix of polymer chains held together by permanent cross-links. The rapid fluctuations of the network chains in the solvent exert an osmotic swelling pressure on the matrix, causing it to expand and generate differences in concentration from one region to another. These large scale variations are static, or change much more slowly than the fluctuations of the network chains. In physical gels such as agarose the network is composed of thick rigid bundles of fibre that cause intense static scattering, but do not fluctuate. Guest polymers contained inside the agarose matrix produce additional scattering. The two sources respectively give rise to a dynamic component $I_{dyn}$ and a static, or pseudo-static component $I_{stat}$ in the total scattered light, such that

$$I = I_{dyn} + I_{stat} \qquad (6)$$

The fraction of light scattered dynamically is then[7-10]

$$X = I_{dyn}/I \qquad (7)$$

When the light scattered by a gel is projected on a screen, $I_{stat}$ appears as an irregular static pattern of dark and bright speckles, while the superimposed fluctuations of $I_{dyn}$ are generally so fast that the human eye cannot detect them. Joosten, McCarthy and Pusey[10] showed that the static speckles obey Poisson statistics, and accordingly have a maximum probability at $I_{stat}=0$. At such positions, therefore, $I=I_{dyn}$. **Figure 1** illustrates the speckle pattern of the total light $I(q)$ scattered by a) a polyacrylamide hydrogel of concentration 100 g/l, and b) a 3 g/l agarose hydrogel containing 7.5 g/l dextran of molecular weight M=610 kDa. Here, the total intensity $I(q)$ is displayed as a function of scattering angle θ in steps of 0.01° in the range 86°≤θ≤91°. According to the Poisson statistics of the static intensity $I_{stat}(q)$, the condition of minimum intensity $I(q)=I_{dyn}$ should be the most frequent.[10] Inspection of Figure 1a, however, shows that for the polyacrylamide sample this expectation is not fulfilled: among the 40 or so minima in the figure the intensity falls below 30 kHz only about 11 times. The lowest measured intensity (*I*=28.4 kHz) is represented by the dashed horizontal line. For the agarose gel in Figure 1b, the discrepancy is even more striking: no clear minimum intensity emerges. For these gels, if a minimum value of *I* does exist, it is attained rather rarely. In these samples, therefore, and contrary to expectation, measurement of the minimum static light scattering intensity does not yield the thermodynamic quantity $I_{dyn}$. The origin of this deviant behaviour, namely depolarised light scattered by the gel, will be discussed in the next section.

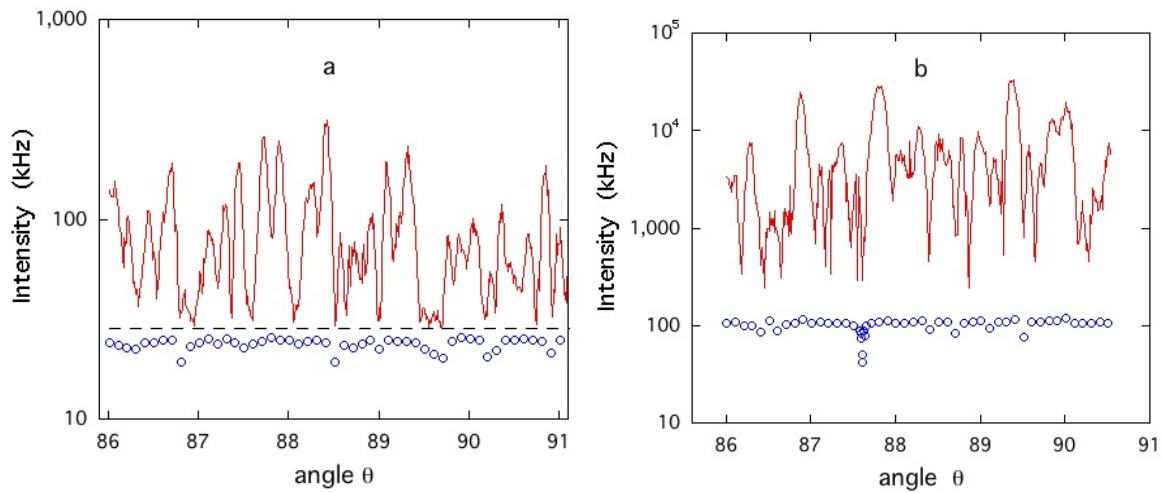

**Figure 1.** Angular dependence of intensity scattered by a) a 100 g/l polyacrylamide hydrogel without polariser, b) a 5 g/l agarose hydrogel containing 7.5 g/l dextran, with polariser. Continuous red lines are the total intensity $I(q)$, open blue symbols are $I_{dyn}(q)$ calculated with Eq. 9.

**Figure 2** illustrates the effect of the static interference pattern on the intensity correlation function $G(q,\tau)$. The inset of the figure shows $G(q,\tau)-1$ for the light scattered by a polyacrylamide hydrogel. The two data sets were obtained at the same scattering angle θ=90°, but for two different sample positions. In one position (blue symbols) $I_{stat}(q)$ is fairly weak, and consequently $G(q,0)-1$ is (≈ 0.8) is large, while in the neighbouring position, where the speckle intensity $I_{stat}(q)$ is strong, $G(q,0)-1$ is small (≈0.08). Since the electric fields of the light scattered by the static inhomogeneities and that of the dynamic fluctuations are coherent in phase, the former acts as a local oscillator at the detector. Eq. 4 then becomes

$$G(\tau) = 1 + f(I)\beta[2X(1-X)g(\tau) + X^2 g(\tau)^2] \quad (8)$$

where $f(I)$ is the instrumental correction function defined earlier. Here we denote $2X(1-X)g(\tau)$ as the *heterodyne* term and $X^2 g(\tau)^2$ is the *homodyne* term, as in Eq. 4. Thus, $I_{stat}=0$ when $X=1$ and Eq. 8 reduces to Eq. 4. Depending on the brightness of the particular speckle, the correlation function $G(\tau)$ adopts a form similar to those in the inset of Figure 2. In neither of those cases, however, is $G(0)-1$ close to its homodyne value $\beta=0.97$.

Insertion of the value of $G(0)-1$ into Eq. 8, together with the condition $g(0)=1$, yields the value of $X$, thus allowing Eq 8 to be solved for the field correlation function $g(\tau)$ at all measured delay times $\tau$.[10] The resulting functions $g(\tau)$ belonging the two spectra in the inset of Figure 2 are displayed in the main figure. Within experimental error they are indistinguishable, and their decay rate is identical. The dynamic intensity $I_{dyn}(q)$ is then

$$I_{dyn}(q) = XI(q) \quad (9)$$

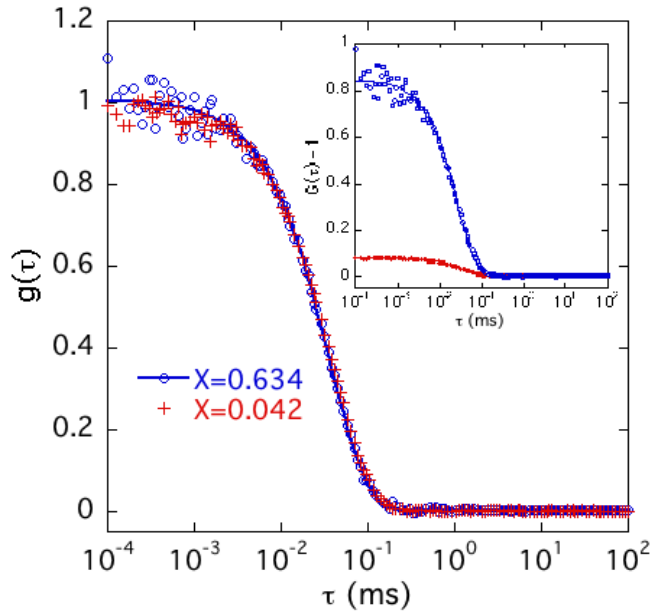

**Figure 2**. Inset: Intensity correlation functions of light scattered at 90º from a poly(acrylamide) hydrogel for two different positions of the sample (O: $X=0.634$, •: $X=0.042$). Main figure: field correlation functions $g(\tau)$ for the same spectra calculated from Eq. 8. Continuous line is least squares fit to a simple exponential decay.

Eq. 9 assumes that the value of $\beta$ that operates on the heterodyne term is identical to that acting on the homodyne term $X^2 g(\tau)^2$, i.e., $\beta_{heterodyne} = \beta$. This assumption is not in general trivial, since the source of the local oscillator intensity, $I_{stat}$, need not coincide with that of the fluctuating light $I_{dyn}$, and the coherence factors are therefore not necessarily identical.[9, 19] To test this assumption, we set $\beta_{heterodyne} = r\beta$, where $r$ may differ from 1. (Note however that with the quasi-single mode optical fibre detection system used here, $r$ is hardly expected to differ significantly from unity.) Then

$$G(\tau)-1=f(I)\beta[2rX(1-X)g(\tau)+X^2 g(\tau)^2] \qquad (10)$$

and hence

$$X = \frac{r}{(2r-1)}\left[1-\left(1-\frac{(G(0)-1)r^2}{(2r-1)f(I)\beta}\right)^{1/2}\right] \qquad (11)$$

To determine the effect of $r$, a set of measurements of $I_{dyn}$ was made at $\theta=90º$, each time changing the position of the sample in the beam, thus shifting the speckle pattern and yielding a wide range of values of $X$. The resulting mean value of $I_{dyn}$ is displayed as a function of $r$ in **Figure 3**. The inset of the figure shows the dependence on $r$ of the normalised variance $(\Delta I_{dyn}^2)^{1/2}/I_{dyn}$. This procedure was performed for two gel systems, i) a 10% poly(acrylamide) hydrogel, in which the network chains are flexible, and ii) a 5 g/l agarose hydrogel containing 7.5 g/l of dextran. Since pure agarose gels without dextran display no measurable dynamic intensity, we assume that the dynamic component stems exclusively from the mobile dextran molecules.

Figure 3 shows that the value of $I_{dyn}$ depends sensitively on $r$. Knowledge of $r$ is therefore essential for correct normalisation in Eq. 2. Although previous DLS observations on different gels or gel-like systems based on the assumption that $r=1$ have shown quantitative agreement with independent measurements of the osmotic pressure[20-22], the present measurements constitute an independent means of determining the value of $r$. The inset in Figure 3 shows that the extremum value of the normalized standard deviation $(\Delta I_{dyn}^2)^{1/2}/I_{dyn}$ is located at $r=1.00$. A similar analysis for the dextran/agarose system also yields $r=1.00$ for the optimum value. (To explore the generality of this conclusion, however, other detection configurations should be investigated, notably with pinhole geometry.) With the present instrument, therefore, these data indicate that the optical coherence factor $\beta$ acting on the heterodyne term is identical to that acting on the homodyne term.

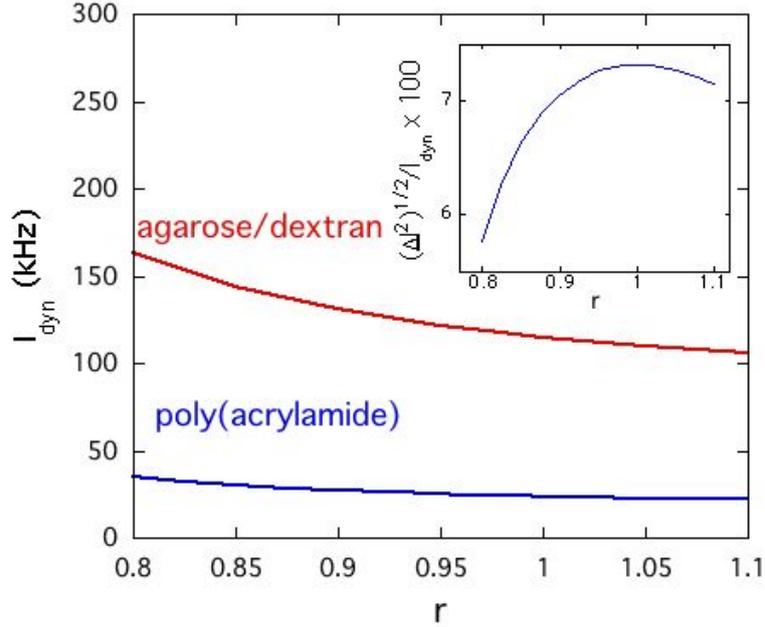

**Figure 3**. Dependence on *r* of the mean value of $I_{dyn}$ (20 measurements) in the 100 g/l polyacrylamide hydrogel (lower curve) and for 7.5 g/l dextran in a 5 g/l agarose gel (upper curve) at 90°. Inset: variation of the normalised standard error $(\Delta I_{dyn}^2)^{1/2}/I_{dyn}$ for polyacrylamide. The location of the extremum at $r=1.00$ validates the assumption that $\beta_{heterodyne} = \beta$ for this system.

*Depolarized scattering*

In the following, we take $r=1$, and turn our attention back to Figure 1 and Eq. 8. The values of $I_{dyn}(q)$ for the polyacrylamide sample displayed in Figure 1a (open blue symbols) are the solutions of Eqs. 8 and 9, expressed in the same kHz units as the total intensity $I(q)$. These results display a high level of apparent noise. Closer inspection of the figure, however, reveals that $I_{dyn}(q)$ is systematically low whenever $I(q)$ is small, i.e., when $I_{stat}$ is close to zero. This behaviour is the consequence of depolarised light scattered by the random network: its horizontal polarization prevents the depolarised electric field from acting as a local oscillator, and it behaves as if it were position-dependent noise. It can be removed either by placing a vertical polariser before the detector (not shown), or more simply by ignoring correlation functions for which *X* is greater than about 0.25. Comparison of the calculated values of $I_{dyn}(q)$ with the minimum values of <*I*> in Figure 1a indicates that the depolarised intensity is approximately 5-10 kHz in this sample.

The case of the agarose hydrogel containing 7.5 g/l dextran in Figure 1b is more demanding. With this sample, the depolarised component is stronger. For the data in Figure 1b, therefore, a vertical polariser was placed before the detector, which successfully eliminates the effects of depolarisation. Nonetheless, occasional major deviations from uniformity of $I_{dyn}(q)$ persist. These are due to occasional gross inhomogeneities in the gel that locally attenuate the scattered light entering the detector, as is illustrated by the large decrease in $I_{dyn}$ at θ=87.6º. Because these attenuation effects are extremely local they do not significantly affect the overall sample transmission, which was measured independently using incoherent light. Such sample defects can be avoided simply by making measurements of $I_{dyn}(q)$ at several speckle positions.

A further remark about the results in Figure 1 is in order. The discrepancy between the minimum value of the total light intensity $I$ and $I_{dyn}$ in Figure 1a, and which is even greater in Figure 1b, is the consequence of multiple scattering. Multiply scattered light is diffuse, and its intensity varies according to the position of the beam in the sample. Figure 1b indicates that, in this particular sample, the multiply scattered light is at least as intense as the polarised light from the mobile guest polymer. By increasing the total luminosity, the contrast between the bright and dark speckles of the singly scattered light is reduced, with the result that $I_{min} > I_{dyn}$. If the local shortfalls in the value of $I_{dyn}$ are overlooked, Figure 1b is proof that, in spite of the difference in optical path between the singly and multiply scattered light, the phase coherence of the multiply scattered light is preserved and the signal remains fully heterodyned. This conclusion is consistent with the long coherence length of light from HeNe laser sources, generally of the order of a metre.

*Osmotic modulus of gels*

The values of $I_{dyn}(q)$ can now be normalised with respect to a scattering standard. With toluene, for example, data are expressed in terms of the Rayleigh ratio at scattering angle θ, $R_θ=I_{dyn}R_{tol}/I_{tol}$, where $R_{tol}=1.35×10^{-7}$ cm$^{-1}$ is the Rayleigh ratio for toluene at λ=6.328×10$^{-5}$ cm. In terms of the normalised $I_{dyn}(q)$, this yields the following expression for the osmotic modulus of a polymer gel

$$κ=c∂Π/∂c=KkTc^2/R_θ \qquad (12)$$

Unlike with static light scattering, there is no need to subtract the signal of the solvent from that of the solution: the solvent response is confined to times much shorter than the relaxation rate of the polymer, and the time discrimination procedure of DLS does not detect it. It is important to note, however, that with solvents other than water, for which the short time response is not necessarily negligible, this background scattering component can affect the apparent value of the optical coherence factor β, and a more elaborate procedure is required to evaluate the parameter *X* in Eq. 8.

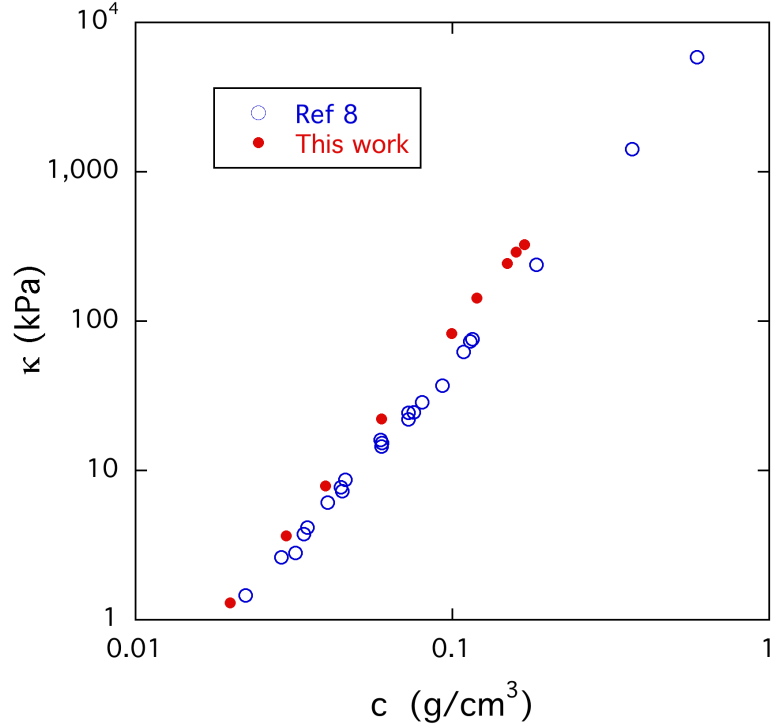

**Figure 4**. Concentration dependence of the osmotic modulus $\kappa=c\partial\Pi/\partial c$ in polyacrylamide gels derived from Eqs. 8 and 9 (filled symbols), compared with previous results from DLS on similar polyacrylamide hydrogels made with pinhole detection geometry.[8]

Formally, the modulus governing the intensity of light scattered by a gel in Eq. 12 is the longitudinal osmotic modulus $M_{os}=\kappa+4G/3$, where *G* is the elastic modulus of the network.[5, 20] Unless the gel is fully swollen in the solvent, however, the elastic term is small and may be neglected. **Figure 4** shows the values of the osmotic modulus κ measured in this way for a set of polyacrylamide gels as a function of polymer concentration, compared with earlier DLS measurements on a similar series of polyacrylamide hydrogels, in which the optical detection was based on a pinhole arrangement.[8] Given that the two sets of samples are not identical, agreement between the two sets of measurements is satisfactory.

The characteristic size of the concentration fluctuations that drive the osmotic pressure in the network is the correlation length ξ. DLS observations detect the *hydrodynamic* correlation length $\xi_H$, which is numerically close to ξ and is found from the Stokes Einstein relationship,

$$\xi_H=(kT/6\pi\eta D), \qquad (13)$$

where η is the viscosity of the solvent, and $D$ is the collective diffusion coefficient.[5] For the 100 g/l polyacrylamide gel in Figure 2 the relaxation rate is $\Gamma=Dq^2=(2.66\pm0.1)\times10^4$ s$^{-1}$, and the value of the diffusion coefficient is

$$D=(7.6\pm0.3)\times10^{-7} \text{ cm}^2/\text{s}. \qquad (14)$$

From Eq. 13 this value yields $\xi_H=31.6\pm1.2$ Å. For these light scattering observations in polyacrylamide, therefore, the Rayleigh condition $q\xi_H \ll 1$ holds, and the dynamic intensity $I_{dyn}(q)$ accordingly displays no measurable variation with scattering angle θ.

*Molecular weight of guest polymers imprisoned in a gel*

In dilute solutions of a polymer of mass $M$ at concentration $c$, the osmotic pressure is given by

$$\Pi = kTc/M \qquad (15)$$

which then yields the standard expression [23]

$$Kc/R_\theta = (1/M_w)[1+(qR_G)^2/3+..](1+2M_wA_2c+..) \qquad (16)$$

where $M_w$ is the weight-average molecular weight of the polymer. In the limit of zero angle, this expression becomes

$$Kc/R_{\theta\to 0} = (1/M_w)(1+2M_wA_2c+..) \qquad (17)$$

**Figure 5** compares the DLS measurements of dextran confined in agarose gels with those in free solution, as a function of polymer concentration. The data in these measurements are extrapolated to zero angle. The values of $Kc/R_{\theta\to 0}$ converge at $c=0$, yielding for the weight average molecular weight $M_w=610\pm80$ kDa, and for the second virial coefficient $A_2=(1.76\pm0.23)\times10^{-4}$ ml.mole/g$^2$. For the dextran confined in the gels, the values of $M_w$ are 550±43 kDa and 576±190 kDa in the 5 g/l and in the 10 g/l agarose gels, respectively. Within experimental error, these three estimates of $M_w$ are indistinguishable. The gel environment however substantially modifies the interaction of the polymer with its surroundings. In the gels, the second virial coefficient of the dextran increases to $A_2=(2.2\pm0.4)\times10^{-4}$ ml.mole/g$^2$ at $c_{gel}=5$ g/l and $(3.0\pm1.7)\times10^{-4}$ at $c_{gel}=10$ g/l. These values are consistent with those of ref. 14, where $A_2$ was reported to be $2.5\times10^{-4}$ ml.mole/g$^2$ at $c_{gel}=5$ g/l. More significantly, the third virial coefficient becomes appreciable, with $A_3=(9.1\pm2.6)\times10^{-3}$ ml$^2$.mole/g$^3$ and (0.7±1.0)

×$10^{-2}$ ml$^2$.mole/g$^3$, respectively. Although the uncertainty in the numerical values of $A_3$ is large, the data in Figure 5 cannot be adequately described without this term. Since the concentrations of dextran and of agarose in the gel are comparable, the magnitude of the third virial coefficient may at first sight appear surprising. In fact, however, it reveals the difference in space filling properties of the rod-like structure of agarose and the more localised excluded volume behaviour of dextran molecules.

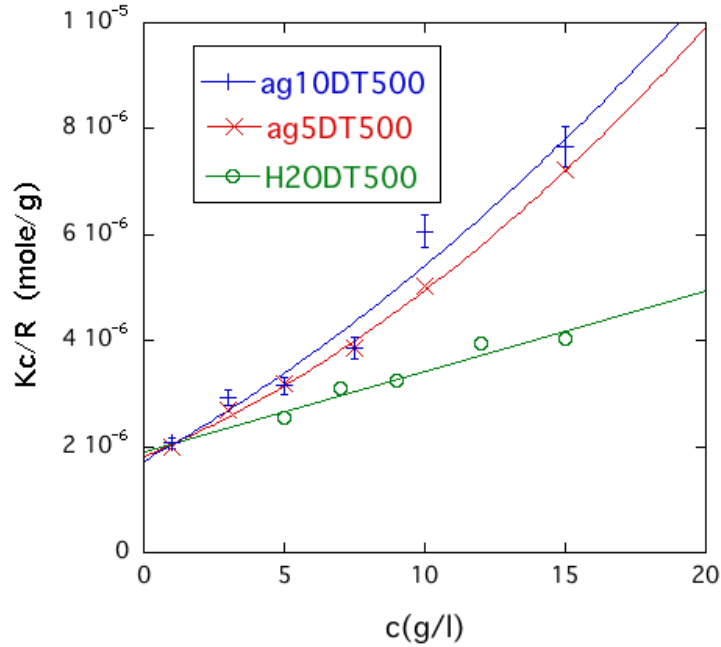

**Figure 5.** Reciprocal of the normalised scattering intensity $R=I_{dyn}(q\to 0)$ from dextran (nominal $M_w$=500 kDa) in free solution (O), and in agarose gels of concentration 5 g/l (×) and 10 g/l (+), as a function of dextran concentration.

We note here that the procedures outlined above for determining $I_{dyn}(q)$ are analogous to that proposed by Joosten *et al.*,[10] whereby the sample is turned successively through a large number of positions, with the individual correlation functions being added. This procedure yields a single intensity correlation function $G(\tau)$ for which $X$=1. Albeit time consuming, the procedure of ref 10 has the advantage of yielding both the average total intensity $<I(q)>$ and the dynamic intensity $I_{dyn}(q)$. The present procedure, by contrast, being interested mainly in $I_{dyn}(q)$, requires only a small number of measurements, from which the effects of depolarized scattering can be eliminated by extrapolating the values of $I_{dyn}(q)$ to $X$=0. The total spatial average of the intensity $<I(q)>$ is found simply by continuously rotating the sample and measuring the average total intensity.

**Conclusions**

This paper draws attention to the fact that, by following certain procedures, DLS experiments yield reliable estimates of the intensity scattered by concentration fluctuations in hydrogels. In gels where light scattered by the osmotically driven fluctuations is heterodyned by that from static or slowly varying inhomogeneities, the two components are separable, yielding good agreement with macroscopic measurements of the osmotic pressure. Gel based DLS intensity measurements of polymer solutions offer appreciable advantages over equivalent measurements in the free solution, notably in removing interference from dust and aggregates. They can also improve the signal to noise ratio at very low concentrations.

Depolarised light scattering, often prevalent in gels, tends to depress the apparent intensity of the osmotic fluctuations, notably in positions of low static speckle intensity. This effect can be circumvented either by placing a polariser before the detector or by restricting the measurements to positions of moderate to high speckle intensity. A procedure is also described whereby the optical coherence factor $\beta$ acting on the homodyne term in the correlation function may be compared to that acting on the heterodyne term. In the present arrangement with a quasi-monomode detection system, the two are shown to be identical. The heterodyne method also yields reliable measurements of light scattered by guest polymers trapped in a hydrogel. The surrounding gel modifies the second and third virial coefficients. The large value of the third virial coefficient in the gel reflects the difference in space filling properties of the rod-like agarose matrix and that of the flexible dextran coil.

**Acknowledgement**

We are indebted to C. Travelet for technical assistance.

**References**

[1] Berne, B.J.; Pecora R. *Dynamic Light Scattering*; Wiley: New York, 1976.
[2] Borsali, R.; Pecora, R. eds. *Soft Matter Characterization*; Springer: New York, 2008.
[3] Chu, B. *Laser Light Scattering*; 2 Ed. Academic Press: San Diego, 1991.
[4] Dusek, K.; Prins, W. *Adv. Polymer Sci*. **1969**, *6,* 1-102.
[5] Tanaka, T.; Hocker, L.O.; Benedek, G.B. *J. Chem Phys*. **1973**, *59*, 5151-5159.
[6] Munch, J.P.; Candau, S.; Duplessix, R.; Picot, C.; Benoit, H. *J. Phys. Lett. (Paris)* **1974**, *35*, L239.

[7] Sellen, D.B. *J. Polymer Sci. Part B: Polymer Phys*. **1987**, *25*, 699-716.

[8] Hecht, A.M.; Geissler, E. *J. Physique (Paris)* **1978**, *39*, 631-638.

[9] Pusey, P.N.; van Megen, W. *Physica A* **1989**, *157*, 705-741.

[10] Joosten, J. G. H.; McCarthy, J. L.; Pusey, P. *Macromolecules* **1991**, *24*, 6690-6699

[11] Hermans, J.J.; Levinson, S. J. *Opt. Soc. Am*. **1951**, *41*, 460-464.

[12] Horkay, F.; Basser, P. J.; Hecht, A.-M.; Geissler, E. *Macromolecules* **2012**, *45*, 2882–2890.

[13] Kloster, C.; Bica, C.; Lartigue, C.; Rochas, C.; Samios, D.; Geissler, E. *Macromolecules* **1998**, *31*, 7712-7716.

[14] Kloster, C.; Bica, C.; Rochas, C.; Samios, D.; Geissler, E. *Macromolecules* **2000**, *33*, 6372-6377.

[15] Morris, C. J. O. R.; Morris, P. *Separation Methods in Biochemistry*; Interscience, Wiley, New York, 1964.

[16] Rochas, C.; Lahaye, M.; Yaphe, W.; Phan Viet, M.T. *Carbohydr. Res*. **1986**, *148*, 199-207.

[17] Rochas, C.; Lahaye, M. *Carbohydr. Polym.***1989**, *10,* 289-298.

[18] Siegert, A.J.F. *MIT Rad. Lab. Rep*. No. 465, 1943.

[19] Joosten. J. G. H.: Geladé, E. T. F.: Pusey, P. N. *Phys. Rev. A* **1990,** *42*, 2161–2175.

[20] Horkay, F.; Burchard, W.; Geissler, E.; Hecht, A.M. *Macromolecules* **1993**, *26*, 1296-1303.

[21] Horkay, F.; Basser, P.J.; Londono, D.J.; Hecht, A-M.; Geissler, E. *J. Chem. Phys*. **2009**, *131*, 184902.

[22] László, K.; Kosik, K.; Rochas, C.; Geissler, E. *Macromolecules* **2003**, *36*, 7771-7776.

[23] Flory, P.J. *Principles of Polymer Chemistry*; Cornell University Press: Ithaca, NY, 1953.

**Table of Contents**

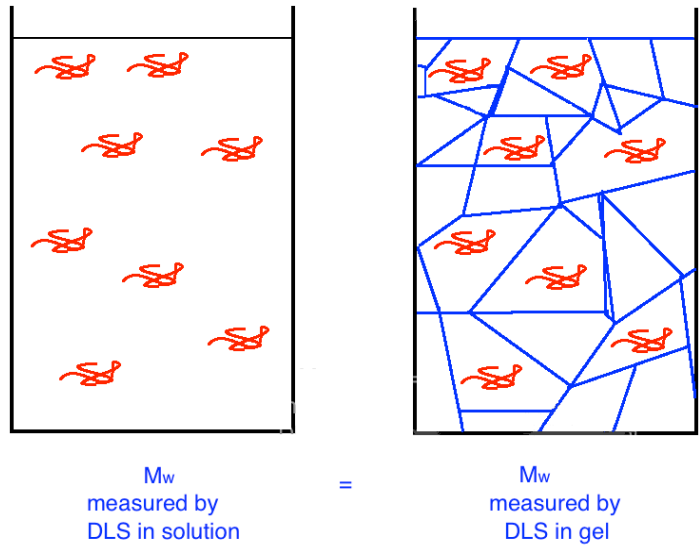

for Table of Contents use only

**Measurement of dynamic light scattering intensity in gels**

Cyrille Rochas and Erik Geissler